\documentclass[a4paper,11pt]{article}

\pdfoutput=1

\usepackage{amsmath,amssymb,mathtools}
\usepackage{color}
\usepackage{graphicx}
\usepackage{subfigure}
\usepackage{cite}
\usepackage[colorlinks=true,linkcolor=red,citecolor=blue,urlcolor=blue,bookmarks]{hyperref}
\usepackage{multirow,makecell}
\usepackage{textcomp}
\usepackage{wasysym}
\usepackage{ulem}

\usepackage{verbatim}

\usepackage[utf8]{inputenc}
\usepackage[T1]{fontenc}

\usepackage[text={16cm,25cm},centering]{geometry} 

\numberwithin{equation}{section}

\def \be {\begin{equation}}
\def \ee {\end{equation}}
\def \ba {\begin{array}}
\def \ea {\end{array}}
\def \bea {\begin{eqnarray}}
\def \eea {\end{eqnarray}}
\def \nn {\nonumber}

\def \a {\alpha}

\def \D {\Delta}

\def \k {\kappa}

\def \s {\sigma}

\def \r {\rho}

\def \cI {\mathcal I}

\def \f {\frac}

\def \inf {\infty}

\def \ep {\mathrm{e}}
\def \ii {\mathrm{i}}

\def \tr {\textrm{tr}}

\def \and {{~\textrm{and}~}}

\def \fer {{\textrm{fer}}}

\def \lf {\lfloor}
\def \rf {\rfloor}
\def \le {\lceil}
\def \re {\rceil}

\begin{document}

\title{
\textbf{Volume-law entanglement fragmentation of quasiparticles}
}
\author{
Jiaju Zhang\footnote{jiajuzhang@tju.edu.cn}
}
\date{}
\maketitle
\vspace{-10mm}
\begin{center}
{\it
Center for Joint Quantum Studiesand Department of Physics, School of Science,\\
Tianjin University, 135 Yaguan Road, Tianjin 300350, China
}
\vspace{10mm}
\end{center}

\begin{abstract}

  We investigate the entanglement entropy in quantum states featuring repeated sequential excitations of unit patterns in momentum space. In the scaling limit, each unit pattern contributes independently and universally to the entanglement entropy, leading to a characteristic volume-law scaling. Crucially, this universal contribution remains identical for both free and interacting models, enabling decomposition of the total entanglement into pattern-specific components. Numerical verification in fermionic and bosonic chains confirms this volume-law fragmentation phenomenon. For fermionic systems, we derive analytical expressions where many-body entanglement becomes expressible through few-body entanglement components. Notably, this analytical framework extends to spin-1/2 XXZ chains through appropriate identifications.

\end{abstract}

\baselineskip 18pt
\thispagestyle{empty}
\newpage


\tableofcontents

\section{Introduction}

Quantum entanglement has been instrumental in various areas of physics, such as quantum information theory, condensed matter physics, and high-energy physics \cite{Amico:2007ag,Eisert:2008ur,Calabrese:2009bph,Laflorencie:2015eck,Rangamani:2016dms,Witten:2018lha}.
Notably, entanglement adheres to distinct scaling laws in various phases of quantum systems, which allows it to serve as an indicator of quantum phase transitions \cite{Osterloh:2002auc,Osborne:2002zz}.
In the context of one-dimensional quantum spin chains with energy gaps, the ground state entanglement entropy typically conforms to the area law. Conversely, in critical spin chains, the ground state entropy adheres to a logarithmic law, with the proportionality constant being related to the central charge \cite{Holzhey:1994we,Vidal:2002rm,Latorre:2003kg,Korepin:2004zz,Calabrese:2004eu}.

The study of the time evolution of entanglement entropy following a quantum quench in many-body systems \cite{Calabrese:2005in} has motivated research into entanglement entropy in excited states. This includes both low-energy states \cite{Alba:2009th,Alcaraz:2011tn,Berganza:2011mh,Pizorn:2012aut,Essler:2012rai,Berkovits:2013mii,Storms:2013wzf,Calabrese:2014ntv,%
Molter:2014qsb,Taddia:2016dbm,Castro-Alvaredo:2018dja,Castro-Alvaredo:2018bij,Castro-Alvaredo:2019irt,Castro-Alvaredo:2019lmj,%
Jafarizadeh:2019xxc,You:2020osa,Zhang:2020vtc,Zhang:2020dtd,Zhang:2021bmy}, where the entropy adheres to either the area law or logarithmic law, and high-energy states \cite{Deutsch:2012vkm,Santos:2012rss,Beugeling:2015plv,Garrison:2015lva,Nakagawa:2017yiw,Lu:2017tbo,Miao:2019xpp,Mohapatra:2024jrb}, where the entropy typically aligns with the volume law. In \cite{Vidmar:2017uux,Vidmar:2017pak,Vidmar:2018rqk,Huang:2019dxk,Lydzba:2020qfx,Bianchi:2021aui}, the average entanglement entropy across the spectrum has been employed to characterize the universal properties of chaotic and integrable systems.

Entanglement of excited states in integrable models can often be understood in terms of quasiparticles \cite{Calabrese:2005in,Alba:2017hlc,Alba:2017lvc}. When a finite number of quasiparticles with large energy and significant momentum differences are excited, a semiclassical picture can describe the entanglement entropy \cite{Castro-Alvaredo:2018dja,Castro-Alvaredo:2018bij,Castro-Alvaredo:2019irt,Castro-Alvaredo:2019lmj}. For earlier related works, see \cite{Pizorn:2012aut,Berkovits:2013mii,Molter:2014qsb}.
On the other hand, when the momentum differences between the excited quasiparticles are small, there are strong coherence effects among the quasiparticles, which results in significant corrections to the semiclassical picture of the entanglement entropy \cite{Zhang:2020vtc,Zhang:2020dtd,Zhang:2021bmy}.

In this paper, we study the entanglement entropy in states with a large number of quasiparticles, where certain unit patterns are excited sequentially and repeatedly in the momentum space. We consider a subsystem consisting of a consecutive block within circular quantum systems, including both free and interacting fermionic and bosonic chains, as well as spin-1/2 XXZ chains. Generally, the entanglement entropy in such states follows a volume law. In \cite{Alba:2009th}, analytical expressions were obtained for fermionic chains in cases where the subsystem size is much smaller than the size of the entire system. In the case of free fermionic chains, we obtain analytical expressions for subsystems of arbitrary sizes, which, in the scaling limit, also apply to interacting fermionic chains and spin-1/2 XXZ chains.

We discover that the coherence among quasiparticles enables the entanglement entropy to be decomposed into separate contributions from distinct parts, each comprising a few sites in the coordinate space and the corresponding unit pattern quasiparticles in the momentum space. In essence, the volume-law entanglement entropy breaks down into components that correspond to the quasiparticle entanglement entropy of much smaller systems. This universal phenomenon of volume-law entanglement fragmentation could be useful for establishing robust entanglement, which is crucial for applications in quantum information processing and quantum computation.

In the remaining part of the paper, we consider free and interacting fermions in sections~\ref{secfreefer} and \ref{secinterfer}, spin-1/2 XXZ chain in section~\ref{secXXZ}, and free and interacting bosons in sections~\ref{secfreebos} and \ref{secinterbos}. We conclude with discussions in section~\ref{secdis}.

\section{Free fermions} \label{secfreefer}

A chain of $L$ free fermions has the Hamiltonian
\be \label{Hfree}
H = \sum_{j=1}^L \Big( a_j^\dag a_j - \frac{1}{2} \Big),
\ee
with the fermionic modes $a_j$ and $a_j^\dag$ satisfying 
$\{a_{j_1},a_{j_2}\}=\{a^\dag_{j_1},a^\dag_{j_2}\}=0$ and $\{a_{j_1},a^\dag_{j_2}\}=\delta_{j_1j_2}$.
The Hamiltonian is already in diagonal form, so we can derive the following analytical expressions.
The excited states with translational invariance are generated by the global modes
\be
b_k^\dag = \frac{1}{\sqrt{L}} \sum_{j=1}^L a_j^\dag \ep^{\frac{2\pi \ii j k}{L}}, ~~
b_k = \frac{1}{\sqrt{L}} \sum_{j=1}^L a_j \ep^{-\frac{2\pi \ii j k}{L}},
\ee
with the momentum $k = 0, 1, \ldots, L-1$. We use the set of excited modes $K = \{k_1, k_2, \ldots\}$ to denote the corresponding excited energy eigenstate $|K\rangle$.

For a subsystem $A$ of $L_A$ consecutive sites in a chain of total length $L$ in state $|K\rangle$, one has the reduced density matrix (RDM) $\rho_{A,K} = \tr_{\bar{A}} (|K\rangle \langle K|)$, from which one gets the R\'enyi entropy and entanglement entropy
\begin{align}
& S_{L,L_A,K}^{(n)} = - \frac{1}{n-1} \log \tr_A (\rho_{A,K}^n), \nonumber \\
& S_{L,L_A,K} = - \tr_A (\rho_{A,K} \log \rho_{A,K}).
\end{align}
The entanglement entropy is just the $n \to 1$ limit of the R\'enyi entropy $S_{L,L_A,K}^{(1)} = S_{L,L_A,K}$.

In the free fermionic chain, one has \cite{Cheong:2002ukf,Vidal:2002rm,Peschel:2002jhw,Latorre:2003kg}
\begin{align}
& S_{L,L_A,K}^{(n),\text{fer}} = - \frac{1}{n-1} \log \det[ C_{L,L_A,K}^n + (1 - C_{L,L_A,K})^n ], \nonumber \\
& S_{L,L_A,K}^{\text{fer}} = \tr_A [ - C_{L,L_A,K} \log C_{L,L_A,K} 
- (1 - C_{L,L_A,K}) \log (1 - C_{L,L_A,K}) ],
\end{align}
which applies to any Hamiltonian for which the number of excitations is conserved, with the Hamiltonian (\ref{Hfree}) as a specific example.
The $L_A \times L_A$ matrix $C_{L,L_A,K}$ has entries
\begin{equation}
[C_{L,L_A,K}]_{j_1j_2} = \langle a_{j_1}^\dag a_{j_2} \rangle_K = h_{L,j_2-j_1,K},
\end{equation}
with $j_1, j_2 = 1, 2, \cdots, L_A$ and the correlation function
\begin{equation} \label{hLjK}
h_{L,j,K} = \frac{1}{L} \sum_{k \in K} \ep^{\frac{2\pi \ii j k}{L}}.
\end{equation}
Note that $S_{L,0,K}^{(n),\text{fer}} = S_{L,L,K}^{(n),\text{fer}} = 0$.

\subsection{Fully occupied states}

We first consider the case that $K$ is composed of the repetition of a unit pattern in the whole momentum space. The unit pattern has $l$ consecutive sites in the momentum space, in which the excited modes are $\k$. The whole system has $L=pl$ sites, and we denote the excited modes of the whole momentum space as $K$ and use the shorthand $K = p\k$.
Explicitly, we have $K = \bigcup_{a=0}^{p-1}( \k + a l )$.
For instance, there are states
\bea
&& l=2, ~ \k=\{0\}, ~ p=4: ~ K=\{ 0 , 2, 4, 6 \}, \nn\\
&& l=3, ~ \k=\{0,1\}, ~ p=2: ~ K=\{0,1,3,4\},
\eea
which can be denoted schematically as
\be
\bullet\!\circ\!\bullet\!\circ\!\bullet\!\circ\!\bullet\!\circ, ~~
\bullet\!\bullet\!\circ\!\bullet\!\bullet\!\circ,
\ee
with the filled circles denoting the excited modes and the empty circles denoting the modes that are not excited.

From (\ref{hLjK}) we get the correlation function
\be
h_{pl,j,p\k} = \delta_{j\,{\rm mod}\,p,0} h_{l,j/p,\k},
\ee
which is possibly nonvanishing only when $j$ is an integer multiple of $p$.
The $L \times L$ correlation matrix of the whole system $C_{L,L,K}$ can be written as a matrix with $l \times l$ blocks, with each block being proportional to the $p \times p$ identity matrix.
The $L_A \times L_A$ correlation matrix $C_{L,L_A,K}$ for the $L_A$-sized subsystem is just the first $L_A$ rows and the first $L_A$ columns of $C_{L,L,K}$.

\begin{figure}[ht]
  \centering
  \includegraphics[width=0.4\textwidth]{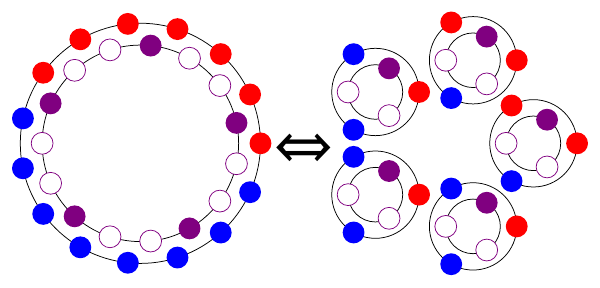}\\
  \caption{Example scheme of the entanglement fragmentation. Each pair of concentric circles symbolizes a system. The larger circle represents the spatial configuration, where red and blue points denote the complementary subsystems. The smaller circle depicts the momentum configuration, with filled purple points indicating excited modes and hollow points representing unexcited modes. For this illustration, we have set the parameters to $L=15$, $L_A=7$, $\alpha=1$, $a=2$, $p=5$, $l=3$, and $\k=\{0\}$.}\label{FigureFragmentation}
\end{figure}

We parameterize the subsystem length as $L_A=\a p+a$, with $\a=0,1,\cdots,l-1$ and $a=0,1,\cdots,p-1$, and the corresponding correlation matrix $C_{L,L_A,K}$ can be transformed to the following form via similarity transformation
\be
C_{pl,\a p+a,p\k} \sim
\underbrace{
C_{l,\a+1,\k}\oplus\cdots\oplus C_{l,\a+1,\k}}_{a}
\oplus
\underbrace{
C_{l,\a,\k}\oplus\cdots\oplus C_{l,\a,\k}}_{p-a}.
\ee
This indicates that the RDM could be written as
\be \label{rplappapk}
\r_{pl,\a p+a,p\k} \sim
\underbrace{
\r_{l,\a+1,\k}\otimes\cdots\otimes\r_{l,\a+1,\k}}_{a}
\otimes
\underbrace{
\r_{l,\a,\k}\otimes\cdots\otimes\r_{l,\a,\k}}_{p-a}.
\ee
There are $p$ copies of the unit pattern $\k$, and each copy is effectively confined in a circle of $l$ sites.
The subsystem has $L_A=\a p+a$ sites, and it effectively decomposes into subsystems across $p$ copies of small systems.
An example scheme of the entanglement fragmentation is shown in Fig.~\ref{FigureFragmentation}.

From the RDM (\ref{rplappapk}), we get the decomposition of the R\'enyi entropy and entanglement entropy
\be
S_{pl,\a p+a,p\k}^{(n),\fer} = a S_{l,\a+1,\k}^{(n),\fer} + (p-a) S_{l,\a,\k}^{(n),\fer}. \label{Spnapapkn}
\ee
Note that we have used the integer $p$ as the number of pattern repetitions.
This inspired us to define the contributions to the entanglement from each unit pattern as
\be \label{snferlkxdef}
s^{(n),\fer}_{l,\k}(x) \equiv 
\f{S^{(n),\fer}_{p l,x p l,p \k}}{p},
\ee
which is the building block for all the results in the paper. Explicitly, we have
\be \label{snferlkx}
s^{(n),\fer}_{l,\k}(x) = y S^{(n),\fer}_{l,\a+1,\k} + ( 1 - y ) S^{(n),\fer}_{l,\a,\k},
\ee
with the parameters
\bea \label{ay}
&& \a \equiv \lfloor l x \rfloor \in \{ 0,1,\cdots,l-1 \}, \nn\\
&& y \equiv (l x{\rm\,mod\,} 1) \in [0,1).
\eea
Here we use $\lfloor l x \rfloor$ to denote the floor function, which gives the largest integer that is smaller than or equal to the argument $l x$.
Note that these formulas are exact, valid for any $p$, $l$, $\a$, $a$, and $\k$.
Furthermore, the function $s_{l,\k}^{(n),\fer}(x)$ is independent of $p$.
For finite $l$ and large $p$, i.e.\ $l\ll p$, the R\'enyi entropy and entanglement entropy (\ref{Spnapapkn}) follow the volume law.
Generally, the formula $s_{l,\k}^{(n),\fer}(x)$ is a piecewise function of $x\in(0,1)$ with $l$ pieces, each piece has length $\f{1}{l}$, and within each piece the function is linear.

We use $|\k|$ to denote the number of excited quasiparticles in the set $\k$. In the special limit, $|\k|\ll l$, the formula (\ref{snferlkx}) becomes
\be \label{limit1}
\lim_{l\to+\inf} s^{(n),\fer}_{l,\k}(x) = \lim_{l\to+\inf} S_{l,xl,\k}^{(n),\fer}.
\ee
In the same limit, the formula (\ref{snferlkxdef}) becomes
\be \label{limit2}
\lim_{l\to+\inf} s^{(n),\fer}_{l,\k}(x) = \lim_{l\to+\inf} \f{S_{pl,xpl,p\k}^{(n),\fer}}{p}.
\ee
The results of (\ref{limit1}) and (\ref{limit2}) should be the same and this is consistent with the fact, found in \cite{Zhang:2020vtc,Zhang:2020dtd,Zhang:2021bmy} that quasiparticles with large momentum differences have independent contributions to the entanglement.

\begin{figure}[ht]
  \centering
  \includegraphics[width=0.66\textwidth]{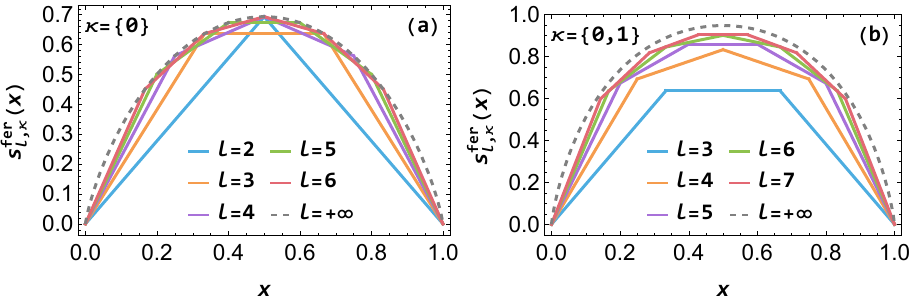}\\
  \caption{The formula (\ref{snferlkx}) of the contribution to the entanglement entropy from each unit pattern in the momentum space of a free fermionic chain.
  In the left panel we use the unit pattern $\k=\{0\}$, and in the right panel we use the unit pattern $\k=\{0,1\}$.
  In each panel, the value of $l$ increases from lower to upper curves.}
  \label{FigureUniversal}
\end{figure}

For $\k=\{0\}$, with $l=2,3,\cdots$, there are \cite{Pizorn:2012aut,Berkovits:2013mii,Molter:2014qsb,Castro-Alvaredo:2018dja,Castro-Alvaredo:2018bij}
\bea
&& \hspace{-3mm} S_{l,l_A,\{0\}}^{(n),\fer} = - \f{1}{n-1} \log \Big[ \Big(\f{l_A}{l}\Big)^n + \Big(1-\f{l_A}{l}\Big)^n  \Big], \nn\\
&& \hspace{-3mm} S_{l,l_A,\{0\}}^{\fer} = - \f{l_A}{l}\log\f{l_A}{l} - \Big(1-\f{l_A}{l}\Big)\log\Big(1-\f{l_A}{l}\Big),
\eea
with which the formula (\ref{Spnapapkn}) is the same as the comb entropy defined in \cite{Keating:2006psq} after one takes the position-momentum duality \cite{Lee:2014nra,Carrasco:2017eul}.
For $\k=\{k_1,k_2\}$ with $l=3,4,\cdots$, the analytical formulas for $S_{l,l_A,\k}^{(n),\fer}$ can be found in \cite{Zhang:2020vtc,Zhang:2020dtd,Zhang:2021bmy}.
Examples of the expression (\ref{snferlkx}) for the entanglement entropy from each unit pattern are shown in Fig.~\ref{FigureUniversal}.

\subsection{Partially occupied states}

In the excited states, the unit pattern $\k$ of length $l$ can be repeated $p$ times in the momentum space and these repeated patterns only occupy a finite ratio of the whole momentum space, leaving the rest of the momentum space unoccupied. We call such state partially occupied states. For example, we may have
\bea
&& l=2, ~ \k=\{0\}, ~ p=2, ~ L=12: ~ K=\{ 0 , 2 \},\\
&& l=3, ~ \k=\{0,1\}, ~ p=2, ~ L=12: ~ K=\{0,1,3,4\},\nn
\eea
which may be denoted as
\be
\bullet\!\circ\!\bullet\!\circ\!\circ\!\circ\!\circ\!\circ\!\circ\!\circ\!\circ\!\circ, ~~
\bullet\!\bullet\!\circ\!\bullet\!\bullet\!\circ\!\circ\!\circ\!\circ\!\circ\!\circ\!\circ.
\ee

\begin{figure}[ht]
  \centering
  \includegraphics[width=0.66\textwidth]{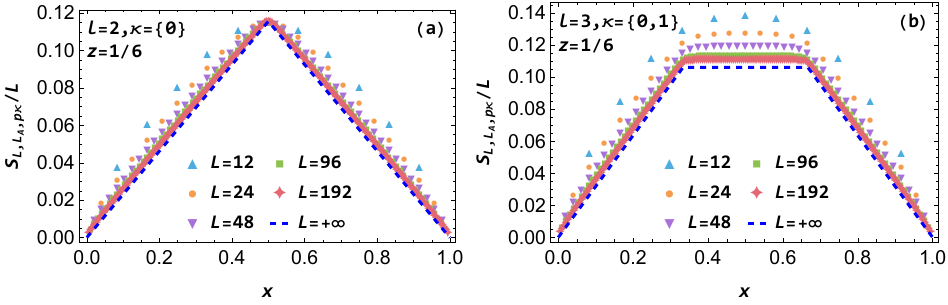}\\
  \caption{The many-body entanglement entropy in the states with partial occupation (symbols) and the result from the few-body entanglement entropy (blue dashed lines).
  On the left panel we use the unit pattern $l=2,\k=\{0\}$, and on the right panel we use the unit pattern $l=3,\k=\{0,1\}$. Note the definition of the repetition ratio $z$ in (\ref{zzz}).}
  \label{FigureFreeFermionsPartial}
\end{figure}

In the scaling limit, we conjecture that each unit pattern continues to contribute independently to the entanglement entropy, and there is
\be
S^{(n),\fer}_{L,x L,p \k} \approx p s^{(n),\fer}_{l,\k}(x),
\ee
with the universal function $s^{(n),\fer}_{l,\k}(x)$ being defined in (\ref{snferlkxdef}) and taking the form (\ref{snferlkx}).
We further get
\be
\label{theformulasIII}
\lim_{L\to+\inf} \f{S_{L,x L,p \k}^{(n),\fer}}{L} = z s_{l,\k}^{(n),\fer}(x), 
\ee
with the definition of the repetition ratio
\be \label{zzz}
z \equiv \lim_{L\to+\inf} \f{p }{L} \in \Big[0,\f1l\Big].
\ee
It is easy to confirm the formula (\ref{theformulasIII}) numerically. We show the examples in Fig.~\ref{FigureFreeFermionsPartial}.

The results indicate the effective form of the RDM
\be
\r_{L,L_A,p\k} \sim
\underbrace{\r_{l,\a+1,\k}\otimes\cdots\otimes\r_{l,\a+1,\k}}_{\lf z a \rf}
\otimes
\underbrace{\r_{l,\a,\k}\otimes\cdots\otimes\r_{l,\a,\k}}_{\lf z ( \lf \f{L}{l} \rf - a) \rf}
\otimes
\r'_{\le (1-z)L_A \re},
\ee
where $\a\equiv\lf x l \rf$,$a\equiv(L_A {\rm~mod~} \lf\f{L}{l}\rf)$, and $\r'_{\le (1-z)L_A \re}$ is a density matrix of about $\le (1-z)L_A \re$ effective sites with sub-volume-law entanglement entropy.

\subsection{States with mixed occupancy}

We also consider more general states, in which different unit patterns are excited in different parts of the momentum space. The parts with different unit patterns may be adjacent or disjoint in the momentum space, and all the excited parts together may or may not occupy the entire spectrum.
Since the relative positions of the different patterns are not important, we denote such states shorthand as $K = \bigcup_{i=1}^r (p_i \k_i)$. Each excited part is characterized by the length of the unit pattern $ l_i $, the mode of the unit pattern $ \k_i $, and the number of repetitions $ p_i $. In the scaling limit, we define
\be \label{zi}
z_i \equiv \lim_{L \to +\infty} \frac{p_i}{L}.
\ee
Of course, there is $\sum_{i=1}^r z_i l_i \in (0,1]$.

For example, we may have the unit pattern with $l_1 = 2$ and $\k_1 = \{0\}$ excited in the momentum space $[0, \frac{L}{3} - 1]$, and the unit pattern with $l_2 = 3$ and $\k_2 = \{0, 1\}$ excited in the momentum space $[\frac{L}{2}, \frac{3L}{4} - 1]$. For $L = 24$, such a state has the excited modes
\be
K = (3\k_1) \bigcup (2\k_2) = \{ 0,2,4,6,12,13,15,16 \},
\ee
and it can be denoted as
\be
\bullet\!\circ\!\bullet\!\circ\!\bullet\!\circ\!\bullet\!\circ\!\circ\!\circ\!\circ\!\circ\!\bullet\!\bullet\!\circ\!\bullet\!\bullet\!\circ\!\circ\!\circ\!\circ\!\circ\!\circ\!\circ.
\ee

\begin{figure}[ht]
  \centering
  \includegraphics[width=0.66\textwidth]{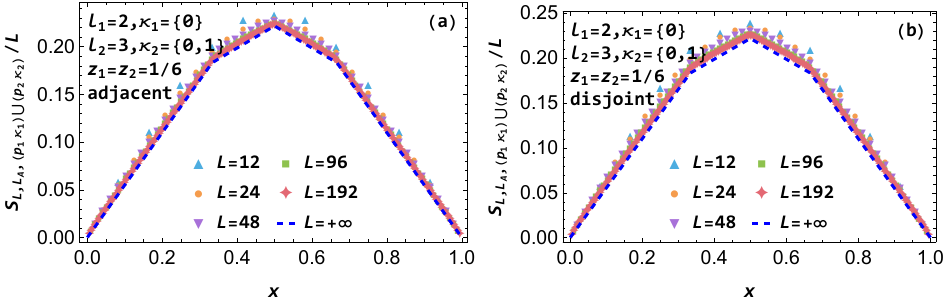}\\
  \caption{The many-body entanglement entropy in the states with partial occupation (symbols) and the result from the few-body   entanglement entropy (blue dashed lines).
  The unit patterns $l_1,\k_1$, $l_2,\k_2$ and their corresponding repetition ratios $z_1$, $z_2$ are shown in each panel.
  On the left panel the two different patterns are adjacent in the momentum space, while on the right panel the two patterns are disjoint.}
  \label{FigureFreeFermionsMixed}
\end{figure}

In such a state with mixed occupancy, we conjecture that each unit pattern still makes an independent and universal contribution to the entanglement entropy
\be
S_{L,xL,\bigcup_{i=1}^r(p_i\k_i)}^{(n),\text{fer}} \approx \sum_{i=1}^r p_i s_{l_i,\k_i}^{(n),\text{fer}}(x),
\ee
which leads to
\be \label{theformulasIV}
\lim_{L \to +\infty} \frac{S_{L,xL,\bigcup_{i=1}^r(p_i\k_i)}^{(n),\text{fer}}}{L} = \sum_{i=1}^r z_i s_{l_i,\k_i}^{(n),\text{fer}}(x).
\ee
This is the main result of this paper. We have expressed the volume-law many-body entanglement entropy as the sum of few-body entanglement entropies. It is easy to confirm the formula (\ref{theformulasIV}) numerically.
The results are also the same as those in the limit $L_A \ll L$ found in \cite{Alba:2009th}.
We show examples of the results in Fig.~\ref{FigureFreeFermionsMixed}.

The results indicate the effective form of the RDM
\be
\r_{L,L_A,\bigcup_{i=1}^r(p_i\k_i)} \sim
\Big[
\bigotimes_{i=1}^r
\Big(
\underbrace{\r_{l_i,\a_i+1,\k_i}\otimes\cdots\otimes\r_{l_i,\a_i+1,\k_i}}_{\lf z_i a_i \rf}
\otimes
\underbrace{\r_{l_i,\a_i,\k_i}\otimes\cdots\otimes\r_{l_i,\a_i,\k_i}}_{\lf z_i ( \lf \f{L}{l_i} \rf - a_i) \rf}
\Big)
\Big]
\otimes
\r'_{\le (1-z)L_A \re},
\ee
where $\a_i\equiv\lf x l_i \rf$,$a_i\equiv(L_A {\rm~mod~} \lf\f{L}{l_i}\rf)$, and $\r'_{\le (1-z)L_A \re}$ is a density matrix of about $\le (1-z)L_A \re$ sites with sub-volume-law entanglement entropy. We have also defined $z=\sum_{i=1}^r z_i$.

\section{Interacting fermions}  \label{secinterfer}

\begin{figure}[ht]
  \centering
  \includegraphics[width=0.66\textwidth]{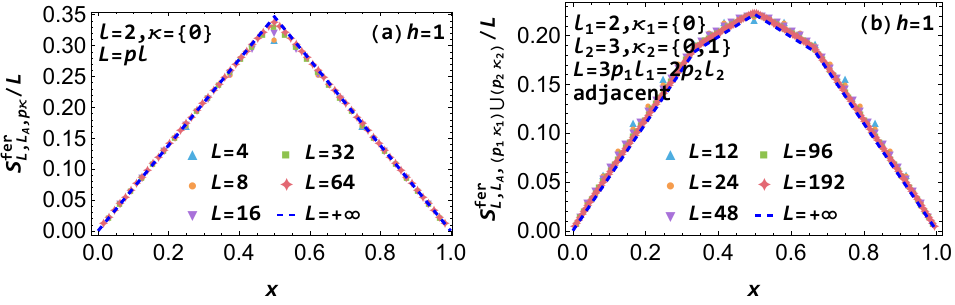}\\
  \caption{The many-body entanglement entropy in interacting fermions (symbols) and the corresponding analytical results from the few-body entanglement entropy (blue dashed lines).
  On the left panel, there is one pattern that occupies the whole momentum space.
  On the right panel, there are two adjacent patterns, with one occupying 1/3 of the momentum space and the other occupying 1/2 of the momentum space.
  We have used $h=1$ for both panels.}
  \label{FigureInteractingFermions}
\end{figure}

The numerical verification of formulas (\ref{theformulasIII}) and (\ref{theformulasIV}) in free fermionic chains motivates us to further conjecture that these formulas still hold true in integrable interacting fermionic chains.
We consider the the transverse field Ising chain
\be
H = -\frac{1}{2} \sum_{j=1}^L \big( \sigma_j^x \sigma_{j+1}^x + h \sigma_j^z \big).
\ee
We use $j=1,2,\cdots,L$ to label the sites of the chain. On each site, there is a qubit represented by Pauli matrices $\sigma_j^x, \sigma_j^y, \sigma_j^z$, and we impose periodic boundary conditions such that $\sigma_{L+1}^x = \sigma_1^x$.
By applying the Jordan-Wigner transformation, the chain is transformed into a fermionic chain with nearest-neighbor interactions.
This model is solvable, as shown in \cite{Lieb:1961fr,Pfeuty:1970ayt}, and the entanglement entropy can be calculated following the methods outlined in \cite{Vidal:2002rm,Latorre:2003kg,Alba:2009th}.
Although we cannot derive the analytical form of the entanglement entropy in such a model, we have carried out extensive numerical verifications of formulas (\ref{theformulasIII}) and (\ref{theformulasIV}) in the Ising chain, thereby confirming the aforementioned conjecture.
We show examples of numerical results in Fig.~\ref{FigureInteractingFermions}.

\section{Spin-1/2 XXZ chain}  \label{secXXZ}

The spin-1/2 XXZ chain has the Hamiltonian
\be
H = -\frac{1}{4} \sum_{j=1}^{L} \big( \sigma_j^x\sigma_{j+1}^x + \sigma_j^y\sigma_{j+1}^y + \Delta\sigma_j^z\sigma_{j+1}^z \big),
\ee
with periodic boundary conditions for the Pauli matrices $\s_{L+1}^\a=\s_1^\a$, $\a=x,y,x$. The excited states in the XXZ chain can be obtained from the coordinate Bethe ansatz \cite{Karbach:1998abi,Karbach:1998oja,Karbach:2000ngx,Caux:2014uuq}.

\begin{figure}[ht]
  \centering
  \includegraphics[width=0.66\textwidth]{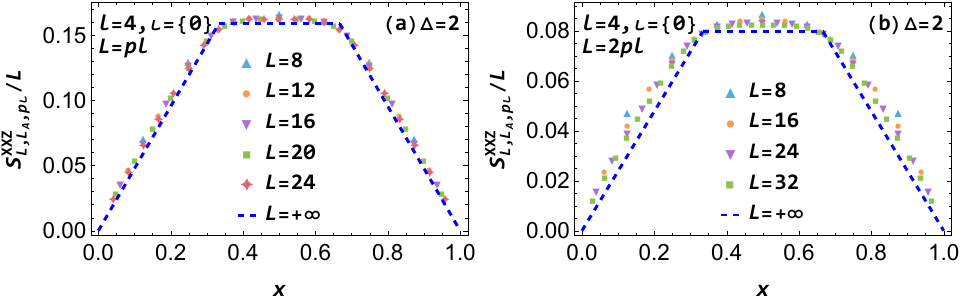}\\
  \caption{Entanglement entropy in states with an extensive number of magnons in the XXZ chain (symbols) is compared with the corresponding analytical formula from the fermionic chain (blue dashed lines).
  On the left panel, there is one pattern that occupies the whole momentum space.
  On the right panel, there is one pattern that occupies 1/2 of the momentum space.
  We have used $\D=2$ for both panels.}
  \label{FigureXXZ}
\end{figure}

We consider the energy eigenstates $|\cI\rangle$ with an extensive number of magnons, which are labeled by the Bethe numbers $\mathcal{I} = \{I_1, I_2, \cdots\}$.
For example, there is a length $ l $ unit pattern $ \iota = \{0\} $ excited in the whole or part of the space of the Bethe numbers $[0, L-1]$, with the number of repetitions being $ p $. When $ \Delta = 2 $, this corresponds to a length $ l-1 $ unit pattern $ \kappa = \{0\} $ with repetition $ p $ in the momentum space. We obtain the R\'enyi entropy and entanglement entropy
\be
S^{(n),\text{XXZ}}_{L,xL,p\iota} \approx p s^{(n),\text{fer}}_{l-1,\kappa}(x),
\ee
which is numerically checked in Fig.~\ref{FigureXXZ}.

\section{Free bosons} \label{secfreebos}

A chain of free bosons has the Hamiltonian
\be
H_{\text{bos}} = \sum_{j=1}^L \left( a_j^\dag a_j + \frac{1}{2} \right),
\ee
with the bosonic modes \(a_j\) and \(a_j^\dag\) satisfying $[a_{j_1}, a_{j_2}] = [a^\dag_{j_1}, a^\dag_{j_2}] = 0$ and $[a_{j_1}, a^\dag_{j_2}] = \delta_{j_1 j_2}$. In an energy eigenstate, the entanglement entropy can be obtained from the subsystem mode method \cite{Zhang:2021bmy}, and the R\'enyi entropy with fixed index $n$ could be obtained from the permanent formula derived in \cite{Zhang:2020dtd} using the wavefunction methods \cite{Bombelli:1986rw,Srednicki:1993im,Castro-Alvaredo:2018dja,Castro-Alvaredo:2018bij}.

\begin{figure}[ht]
  \centering
  \includegraphics[width=0.66\textwidth]{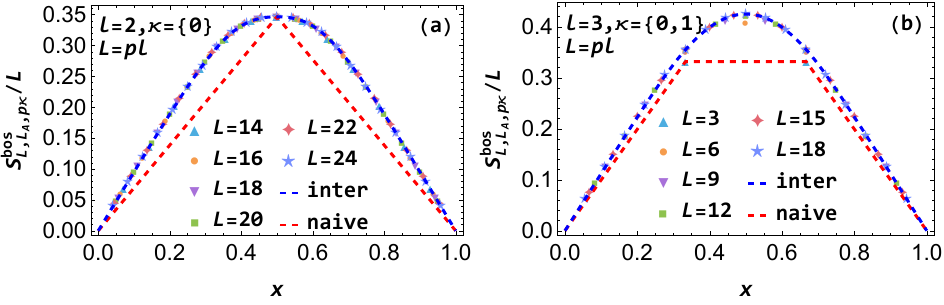}\\
  \caption{The entanglement entropy in free bosonic chains (symbols) is contrasted with the naive expectation $s_{l,\k}^{(n),\text{naive}}(x)$ (\ref{snnaivelkxdef}) (red dashed lines). In each panel, the blue dashed line represents the interpolation of the results corresponding to the maximum $L$, i.e. $L=24$ for panel (a) and $L=18$ for panel (b).
  In each panel, there is one pattern that occupies the whole momentum space.}
  \label{FigureFreeBosons}
\end{figure}

We consider the entanglement entropy in a state with repeated unit patterns in the bosonic chain. Following (\ref{snferlkxdef}) in the fermionic chain, we define the finite contributions to the entanglement from each unit pattern
\be \label{slikinbosx}
s^{(n),\text{bos}}_{l,\k}(x) \equiv \lim_{p \to +\infty} \frac{S_{pl,xpl,p\k}^{(n),\text{bos}}}{p}.
\ee
Following (\ref{snferlkx}), one might naively expect that the function \( s^{(n),\text{bos}}_{l,\k}(x) \) takes the form
\be \label{snnaivelkxdef}
s_{l,\k}^{(n),\text{naive}}(x) \equiv y S_{l,\a+1,\k}^{(n),\text{bos}} + (1 - y) S_{l,\a,\k}^{(n),\text{bos}},
\ee
with the definitions of \(\a\) and \(y\) given in (\ref{ay}).
However, numerical results show that while the function \( s^{(n),\text{bos}}_{l,\k}(x) \) is well-defined, it differs from the naive expectation \( s_{l,\k}^{(n),\text{naive}}(x) \).
We present numerical examples in Fig.~\ref{FigureFreeBosons}.
It remains an open problem to derive the explicit analytical expression of $s^{(n),\text{bos}}_{l,\k}(x)$.
The bosonic version of the formulas (\ref{theformulasIII}) and (\ref{theformulasIV}) still apply with the definitions (\ref{slikinbosx}), and we have verified the formulas with numerically examples, which are not shown here.

\section{Interacting bosons} \label{secinterbos}

\begin{figure}[ht]
  \centering
  \includegraphics[width=0.99\textwidth]{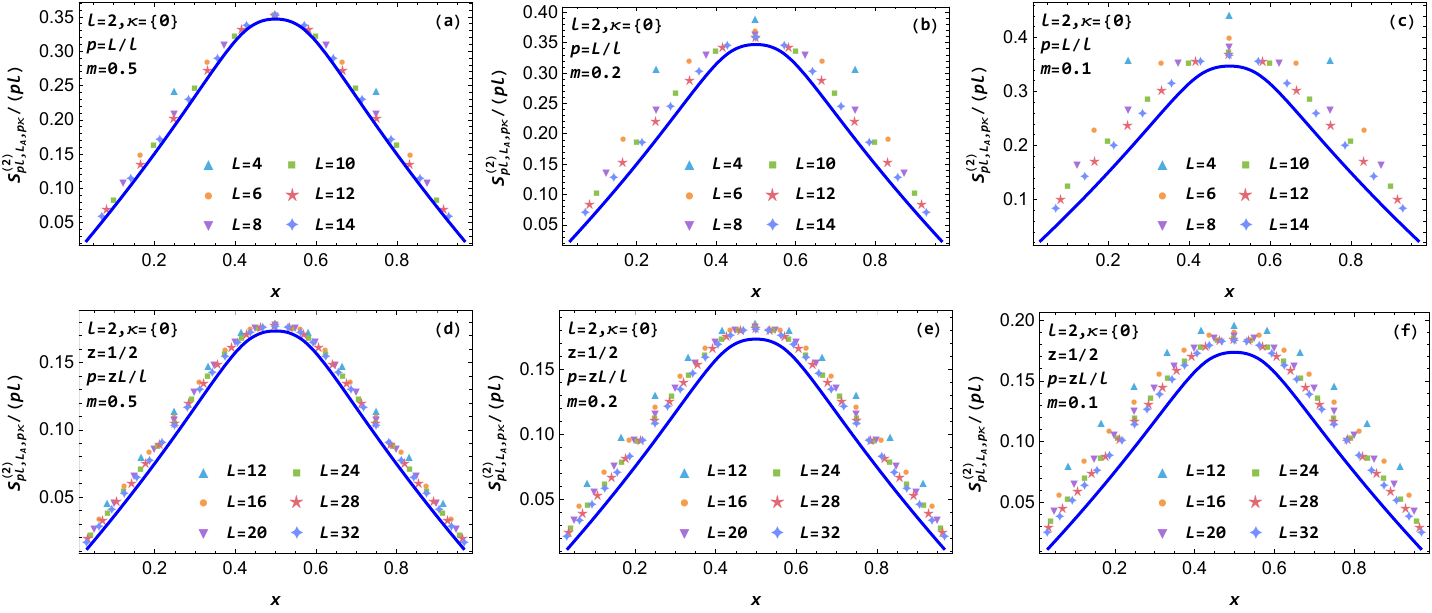}\\
  \caption{The many-body entanglement entropy in interacting bosons (symbols) and the results $s_{l_1,\k_1}^{(2)}(x)$ and $z s_{l_1,\k_1}^{(2)}(x)$ (blue lines) with $l_1=2,\k_1=\{0\}$ and $s_{l,\k}^{(2)}(x)$ taking the approximate value from the interpolation in panel (a) of Fig.~\ref{FigureFreeBosons}.
  On the top panels, there is one pattern that occupies the whole momentum space. On the bottom panels, there is one pattern that occupies 1/2 of the momentum space. In different columns, we have used different values of $m$.}
  \label{FigureInteractingBosons}
\end{figure}

Similarly to (\ref{theformulasIV}) in the fermionic chain, we conjecture that in the free and interacting bosonic chains, there is
\be
\lim_{L \to +\infty} \frac{S_{L,xL,\bigcup_{i=1}^r(p_i\k_i)}^{(n),\text{bos}}}{L} = \sum_{i=1}^r z_i s_{l_i,\k_i}^{(n),\text{bos}}(x),
\ee
with the definition of $z_i$ in (\ref{zi}) and the definition of $s_{l_i,\k_i}^{(n),\text{bos}}(x)$ in (\ref{slikinbosx}).
We consider the harmonic chain
\be
H = \frac{1}{2} \sum_{j=1}^L \left[ p_j^2 + m^2 q_j^2 + (q_j - q_{j+1})^2 \right],
\ee
with the periodic boundary condition $q_{L+1} = q_j$ and the commutation relations $[q_{j_1}, q_{j_2}] = [p_{j_1}, p_{j_2}] = 0$ and $[q_{j_1}, p_{j_2}] = i \delta_{j_1 j_2}$. The harmonic chain is essentially a chain of bosons with nearest-neighbor interactions. The above conjecture could be easily checked by numerically calculating the R\'enyi entropy in excited energy eigenstates following the wavefunction methods \cite{Castro-Alvaredo:2018dja,Castro-Alvaredo:2018bij,Zhang:2020ouz,Zhang:2020txb}.
In Fig.~\ref{FigureInteractingBosons}, we provide numerical results supporting the conjecture.
For larger masses $m$, the model approaches free bosons and exhibits smaller numerical errors. Conversely, smaller masses $m$ lead to larger numerical errors.

\section{Discussions} \label{secdis}

We have studied the entanglement entropy in quasiparticle states that exhibit repeated and sequential unit patterns in momentum space. We found that each unit pattern contributes independently and universally to the entanglement in the scaling limit, leading to a volume-law scaling of entanglement entropy. Numerical simulations confirmed these findings, demonstrating volume-law entanglement fragmentation in both fermionic and bosonic chains. We derived an analytical formula for free fermions that is applicable not only to interacting fermions in the scaling limit but also, with certain identifications, to the spin-1/2 XXZ chain. Numerical results of free and interacting bosonic chains reveal universal entanglement characteristics in bosonic systems; however, explicit analytical formulations remain elusive. Deriving such analytical expressions for bosonic chains would be an important direction for future work.

The results in this paper reveal a fundamental distinction between fragmented volume-law entanglement and thermal volume-law entanglement. Thermal volume-law entanglement, found in states such as random states or eigenstates of chaotic Hamiltonians \cite{Garrison:2015lva,Nakagawa:2017yiw,Bianchi:2021aui}, emerges from maximal scrambling and adherence to the eigenstate thermalization hypothesis (ETH), resulting in featureless, homogeneous entanglement devoid of substructure. In contrast, fragmented volume-law entanglement in quasiparticle states arises from repeated unit patterns in momentum space and decomposes the state into a sum of few-body entangled components. Although both types exhibit extensive entanglement, fragmented states violate ETH due to the underlying integrability, contrasting sharply with the thermal case.
The existence and universality of entanglement fragmentation remain to be further studied. We conjecture that this phenomenon may stem from the existence and independence of quasiparticles resulting from integrability.

It was suggested in \cite{Castro-Alvaredo:2018dja} to harness the entanglement of quasiparticles for quantum information purposes. However, the entanglement among a small number of quasiparticles is delicate and susceptible to environmental disruptions.
The results presented in this paper imply that by successively exciting repeated patterns in momentum space, it is possible to create a robust volume-law entanglement of quasiparticles.
The universality of fragmentation suggests potential for robust entanglement, though stability to environmental perturbations requires further study.

\section*{Acknowledgements}

The author thanks Olalla A. Castro-Alvaredo and M.~A. Rajabpour for reading a previous versions of the draft and helpful comments and discussions. JZ acknowledgements support from the National Natural Science Foundation of China (NSFC) grant number 12205217 and Tianjin University Self-Innovation Fund Extreme Basic Research Project grant number 2025XJ21-0007.

\providecommand{\href}[2]{#2}\begingroup\raggedright\endgroup


\end{document}